\documentclass{article}

\usepackage{amssymb,latexsym,amsmath}

\usepackage[pdftex]{graphicx}

\usepackage{hyperref}

\begin{document}

\pagestyle{plain}

\newtheorem{theorem}{Theorem}[section]

\newtheorem{proposition}[theorem]{Proposition}

\newtheorem{lema}[theorem]{Lemma}

\newtheorem{corollary}[theorem]{Corollary}

\newtheorem{definition}[theorem]{Definition}

\newtheorem{remark}[theorem]{Remark}

\newtheorem{exempl}{Example}[section]

\newenvironment{exemplu}{\begin{exempl}  \em}{\hfill $\square$

\end{exempl}}

\newcommand{\ea}{\mbox{{\bf a}}}

\newcommand{\eu}{\mbox{{\bf u}}}

\newcommand{\ueu}{\underline{\eu}}

\newcommand{\ueo}{\overline{u}}

\newcommand{\oeu}{\overline{\eu}}

\newcommand{\ew}{\mbox{{\bf w}}}

\newcommand{\ef}{\mbox{{\bf f}}}

\newcommand{\eF}{\mbox{{\bf F}}}

\newcommand{\eC}{\mbox{{\bf C}}}

\newcommand{\en}{\mbox{{\bf n}}}

\newcommand{\eT}{\mbox{{\bf T}}}

\newcommand{\eL}{\mbox{{\bf L}}}

\newcommand{\eR}{\mbox{{\bf R}}}

\newcommand{\eV}{\mbox{{\bf V}}}

\newcommand{\eU}{\mbox{{\bf U}}}

\newcommand{\ev}{\mbox{{\bf v}}}

\newcommand{\eve}{\mbox{{\bf e}}}

\newcommand{\uev}{\underline{\ev}}

\newcommand{\eY}{\mbox{{\bf Y}}}

\newcommand{\eK}{\mbox{{\bf K}}}

\newcommand{\eP}{\mbox{{\bf P}}}

\newcommand{\eS}{\mbox{{\bf S}}}

\newcommand{\eJ}{\mbox{{\bf J}}}

\newcommand{\eB}{\mbox{{\bf B}}}

\newcommand{\eH}{\mbox{{\bf H}}}

\newcommand{\leb}{\mathcal{ L}^{n}}

\newcommand{\eI}{\mathcal{ I}}

\newcommand{\eE}{\mathcal{ E}}

\newcommand{\hen}{\mathcal{H}^{n-1}}

\newcommand{\eBV}{\mbox{{\bf BV}}}

\newcommand{\eA}{\mbox{{\bf A}}}

\newcommand{\eSBV}{\mbox{{\bf SBV}}}

\newcommand{\eBD}{\mbox{{\bf BD}}}

\newcommand{\eSBD}{\mbox{{\bf SBD}}}

\newcommand{\ecs}{\mbox{{\bf X}}}

\newcommand{\eg}{\mbox{{\bf g}}}

\newcommand{\paromega}{\partial \Omega}

\newcommand{\gau}{\Gamma_{u}}

\newcommand{\gaf}{\Gamma_{f}}

\newcommand{\sig}{{\bf \sigma}}

\newcommand{\gac}{\Gamma_{\mbox{{\bf c}}}}

\newcommand{\deu}{\dot{\eu}}

\newcommand{\dueu}{\underline{\deu}}

\newcommand{\dev}{\dot{\ev}}

\newcommand{\duev}{\underline{\dev}}

\newcommand{\weak}{\stackrel{w}{\approx}}

\newcommand{\mild}{\stackrel{m}{\approx}}

\newcommand{\lrightarrow}{\stackrel{L}{\rightarrow}}

\newcommand{\rrightarrow}{\stackrel{R}{\rightarrow}}

\newcommand{\strong}{\stackrel{s}{\approx}}

\newcommand{\weakdown}{\rightharpoondown}

\newcommand{\opg}{\stackrel{\mathfrak{g}}{\cdot}}

\newcommand{\opunu}{\stackrel{1}{\cdot}}
\newcommand{\opdoi}{\stackrel{2}{\cdot}}

\newcommand{\opn}{\stackrel{\mathfrak{n}}{\cdot}}
\newcommand{\opx}{\stackrel{x}{\cdot}}

\newcommand{\tr}{\ \mbox{tr}}

\newcommand{\Ad}{\ \mbox{Ad}}

\newcommand{\ad}{\ \mbox{ad}}

\renewcommand{\contentsname}{ }

\title{$\lambda$-Scale, a lambda calculus for spaces with dilations}

\author{Marius Buliga \\ 
\\
Institute of Mathematics, Romanian Academy \\
P.O. BOX 1-764, RO 014700\\
Bucure\c sti, Romania\\
{\footnotesize Marius.Buliga@imar.ro}}

\date{This version: 23.05.2012}

\maketitle

\begin{abstract}
$\lambda$-Scale is an enrichment of lambda calculus  which is adapted to emergent algebras. It can be used therefore in metric spaces with dilations. 
\end{abstract}

\section{Introduction, background}

My goal is to propose an enrichment of $\lambda$-calculus \cite{church} \cite{baren} which contains the formalism of emergent algebras. It can be used in particular for metric spaces with dilations.

$\Gamma$-idempotent right quasigroups have been introduced in \cite{buligairq}. Uniform idempotent right quasigroups are called in that paper "emergent algebras". Here I shall use the name "emergent algebra" for a $\Gamma$-irq. 

Emergent algebras (in the original sense from \cite{buligairq}) have been studied further in the paper \cite{buligabraided}, in relation with metric spaces with dilations introduced in \cite{buligadil1}. See \cite{buligaintro} for an introduction into metric spaces with dilations and their intrinsic approximate differential calculus. 

In the paper \cite{buligachora} I tried to study emergent algebras by using a graphical formalism of decorated tangles, under the banner "computing with space". This paper is an effort to associate a rigorous calculus to that formalism, explained in detailed in sections 1 and 3 to 8 from \cite{buligachora}, which are  recommended for better understanding of the project of computing with space.  

\paragraph{Acknowledgement.} This work was supported by a grant of the Romanian National Authority for Scientific Research, CNCS – UEFISCDI, project number 
PN-II-ID-PCE-2011-3-0383.

\subsection{Emergent algebras, spaces with dilations}

\begin{definition} A right quasigroup is a set $X$ with a binary operation 
$\circ$ such that for each $a, b \in X$ there exists a unique $x \in X$ such that 
$a \, \circ  \, x \, = \, b$. We write the solution of this equation 
$x \, = \, a \, \bullet \, b$. 

An idempotent right quasigroup (irq) is a  right quasigroup $(X,\circ)$ such that 
 for any $x \in X$ $x \, \circ \, x \, = \, x$. Equivalently, it can be seen as a 
  set $X$ endowed with two  operations $\circ$ and $\bullet$, which satisfy the following axioms: for any $x , y \in X$  
\begin{enumerate}
\item[(R1)] \hspace{2.cm} $\displaystyle x \, \circ \, x \, = \, x \, \bullet \, x \,  = \,  x$
\item[(R2)] \hspace{2.cm} $\displaystyle x \, \circ \, \left( x\, \bullet \,  y \right) \, = \, x \, \bullet \, \left( x\, \circ \,  y \right) \, = \, y$
\end{enumerate}
\label{defquasigroup}
\end{definition}

Perhaps the most well known examples of irqs are quandles. They were introduced by 
 Joyce \cite{joyce}, in relation with the  Reidemeister
moves from knot theory.

\begin{definition}
A quandle is a irq which is left self-distributive, i.e. it satisfies the supplementary axiom: 
\begin{enumerate}
\item[(R3)] for any $x,y, z \in X$ and for any choice of operations 
$\displaystyle *_{1}, *_{2} \in \left\{ \circ, \bullet \right\}$ we have 
$\displaystyle x \, *_{1} \left( y \, *_{2} \,  z \right) \, = \, \left( x \, *_{1} y \right) *_{2} \left( x *_{1} z \right)$. 
\end{enumerate}
\label{defquandle}
\end{definition}

Next is the definition of a $\Gamma$-irq, the object called in this paper "emergent algebra". For explanations of this denominations see the paper \cite{buligairq}. 

\begin{definition}
Let $\Gamma$ be a commutative group. A $\Gamma$-idempotent right quasigroup 
is a set $X$ with a function $\displaystyle \varepsilon \in \Gamma \mapsto 
\circ_{\varepsilon}$ such that:
\begin{enumerate}
\item[(a)] for any $\varepsilon \in \Gamma$  $\displaystyle (X, \circ_{\varepsilon})$ is an irq, 
\item[(b)] let $1 \in \Gamma$ be the neutral element; then for any $x, y \in X$ we have 
$\displaystyle x \, \circ_{1} \, y \, = \, y$, 
\item[(c)] for any $\varepsilon, \mu \in \Gamma$ and any $x, y \in X$ we have 
$\displaystyle x \, \circ_{\varepsilon} \, \left( x \, \circ_{\mu} \, y \right) \, = \, 
x \, \circ_{\varepsilon \mu} \, y$. 
\end{enumerate}
\label{defgammairq}
\end{definition}

In the realm of metric spaces, the object corresponding to (the local version of) an emergent algebra is the one of a dilatation structure \cite{buligadil1}. Later, in \cite{vodosel} the authors generalize dilatation structures to quasimetric spaces and introduce the name "quasimetric space with dilations". In \cite{buligairq}, \cite{buligabraided} I proved a general result about the emergence of algebraic and differential structure from a {\em uniform} $\Gamma$-idempotent right quasigroup, which applies to metric case (dilatation structures) or to quasimetric case (quasimetric spaces with dilations). It is shown that is not a pure metric phenomenon, in fact   metric (i.e. distance function) or quasimetric are needed only in order to have an uniform structure over the space. That is why I prefer now the name "space with dilations" instead the initial "dilatation structure". 

Here are, in order, the definition of a uniform  $\Gamma$-idempotent right quasigroup and then the definition of a metric space with dilations. (See \cite{buligabraided} definition 5.2 for uniform idempotent right quasigroups endowed with a class of absolutes.)

\begin{definition}
A $\Gamma$-uniform irq  $(X, \circ)$ is a separable uniform  
space $X$ which is also a $\Gamma$-irq, with continuous operations, endowed with an absolute 
$0$ (topological filter over $\Gamma$ which is translation invariant)  such that: 
\begin{enumerate}
\item[(C)] the operation $\circ$ is compactly contractive: for each compact set 
$K \subset X$ and open set $U \subset X$, with $x \in U$, there is an open set 
$\displaystyle A(K,U) \subset \Gamma$ with $\mu(A) = 1$ for any $\mu \in 
Abs(\Gamma)$ and for any $u \in K$ and
$\varepsilon \in A(K,U)$, we have $\displaystyle x \circ_{\varepsilon} u \in U$; 
\item[(D)] the following limits exist 
$$ \lim_{\varepsilon \rightarrow 0} \Delta_{\varepsilon}^{x}(u,v)  \, = \, 
\Delta^{x}(u,v) \quad , \quad \lim_{\varepsilon \rightarrow 0} \Sigma_{\varepsilon}^{x}(u,v) 
 \,  = \, \Sigma^{x}(u,v) $$
and are uniform with respect to $x, u, v$ in a compact set. 
\end{enumerate}
\label{deftop}
\end{definition}

\begin{definition}
 A metric space with dilations $(X,d, \delta)$ is a triple formed by: 
\begin{enumerate}
\item[-] $(X,d)$ a complete metric space such that for any $x  \in X$ the 
closed ball $\bar{B}(x,3)$ is compact, 
\item[-] an assignment to any $x \in X$  and $\varepsilon \in (0,+\infty)$ 
of a  homeomorphism, defined as: if 
$\displaystyle   \varepsilon \in (0, 1]$ then  $\displaystyle 
 \delta^{x}_{\varepsilon} : U(x)
\rightarrow V_{\varepsilon}(x)$, else 
$\displaystyle  \delta^{x}_{\varepsilon} : 
W_{\varepsilon}(x) \rightarrow U(x)$,
\end{enumerate}
with the following properties.  
\begin{enumerate}
\item[{\bf A0.}]  For any $x \in X$ the sets $ \displaystyle U(x), V_{\varepsilon}(x), 
W_{\varepsilon}(x)$ are open neighbourhoods of $x$.  There are   $1<A<B$ such that for any $x \in X$  and any 
$\varepsilon \in (0,1)$ we have: 
$$\displaystyle  B_{d}(x, \varepsilon) \subset \delta^{x}_{\varepsilon}  B_{d}(x, A) 
\subset V_{\varepsilon}(x) \subset $$ 
$$\subset W_{\varepsilon^{-1}}(x) \subset \delta_{\varepsilon}^{x}  B_{d}(x, B)$$
Moreover for any compact set $K \subset X$ there are $R=R(K) > 0$ and 
$\displaystyle \varepsilon_{0}= \varepsilon(K) \in (0,1)$  such that  
for all $\displaystyle u,v \in \bar{B}_{d}(x,R)$ and all 
$\displaystyle \varepsilon  \in (0,\varepsilon_{0})$,  we have 
$\displaystyle 
\delta_{\varepsilon}^{x} v \in W_{\varepsilon^{-1}}(
\delta^{x}_{\varepsilon}u)$. 

\item[{\bf A1.}]  For any $x \in X$ 
$\displaystyle  \delta^{x}_{\varepsilon} x = x $ and $\displaystyle \delta^{x}_{1} = id$. 
Consider  the closure $\displaystyle Cl(dom \, \delta)$ of the set 
$$ dom \, \delta = \left\{ (\varepsilon, x, y) \in (0,+\infty) \times X 
\times X \mbox{ : } \right.$$ 
$$\left. \mbox{ if } \varepsilon \leq 1 \mbox{ then } y 
\in U(x) \,
\, , \mbox{  else } y \in W_{\varepsilon}(x) \right\} $$ 
seen in  $[0,+\infty) \times X \times X$ endowed with
 the product topology. The function $\displaystyle \delta : dom \, \delta 
\rightarrow  X$,  $\displaystyle \delta (\varepsilon,  x, y)  = 
\delta^{x}_{\varepsilon} y$ is continuous, admits a continuous extension 
over $\displaystyle Cl(dom \, \delta)$ and we have 
$\displaystyle \lim_{\varepsilon\rightarrow 0} \delta_{\varepsilon}^{x} y \, =
\, x$. 

\item[{\bf A2.}] For any  $x, \in X$, $\displaystyle \varepsilon, \mu \in (0,+\infty)$
 and $\displaystyle u \in U(x)$, whenever one of the sides are well defined
   we have the equality 
$\displaystyle  \delta_{\varepsilon}^{x} \delta_{\mu}^{x} u  =
\delta_{\varepsilon \mu}^{x} u$.

\item[{\bf A3.}]  For any $x$ there is a distance  function $\displaystyle (u,v) \mapsto d^{x}(u,v)$, defined for any $u,v$ in the closed ball (in distance d) $\displaystyle 
\bar{B}(x,A)$, such that uniformly with respect to $x$ in compact set we have
the limit: 
$$\lim_{\varepsilon \rightarrow 0} \quad \sup  \left\{  \mid 
\frac{1}{\varepsilon} d(\delta^{x}_{\varepsilon} u, \delta^{x}_{\varepsilon} v) \ - \ d^{x}(u,v) \mid \mbox{ :  } u,v \in \bar{B}_{d}(x,A)\right\} \ =  \ 0$$

\item[{\bf A4.}] Let us define 
$\displaystyle \Delta^{x}_{\varepsilon}(u,v) =
\delta_{\varepsilon^{-1}}^{\delta^{x}_{\varepsilon} u} \delta^{x}_{\varepsilon} v$. 
Then we have the limit, uniformly with respect to $x, u, v$ in compact set,  
$$\lim_{\varepsilon \rightarrow 0}  \Delta^{x}_{\varepsilon}(u,v) =  \Delta^{x}(u, v)  $$
 \end{enumerate}
\label{defweakstrong}
\end{definition}

If we neglect the problems related to the domanins and codomains of dilations, then we remark that $(X,\circ)$, with 
$$\displaystyle x \circ_{\varepsilon} y = \delta^{x}_{\varepsilon} y$$ 
is a uniform $\Gamma$-irq, with $\Gamma = (0,+\infty)$ with multiplication and the absolute $0$ is the topological filter of the real number $0$ restricted to $(0, + \infty)$.

\section{$\lambda$-Scale calculus}

In this section is introduced the $\lambda$-Scale calculus. In the section \ref{secrel} is introduced the relative $\lambda$-Scale calculus. 

This calculus is an enhancement of untyped lambda calculus with $\beta$-reduction, extensionality rules and substitution. It can be seen as lambda calculus with a new "dilation" operation (taken from emergent algebras), but it is interesting to see how the dilation operation and the application operation (from lambda calculus) merge into a "scaled" operation $(A,B) \mapsto A \varepsilon B$, with $\varepsilon$ a parameter in a commutative group $\Gamma$.

\subsection{Terms and operations} 

$X$ is the set of variables. $T$ is the set of terms (or trees). $\lambda$ is the abstraction operation. $\Gamma$ is an abelian group. We use parantheses "$($" and "$)$".

\begin{definition}
Terms are constructed according to the following rules.
\begin{enumerate}
\item[-] variables are terms: $X \subset T$, 
\item[-] if $x \in X$ and $A \in T$ then $x \lambda A \in T$, 
\item[-] if $A, B \in T$ and $\varepsilon \in \Gamma$ then $A \varepsilon B \in T$, 
\item[-] any term is obtained after a finite combination of the previous rules. 
\end{enumerate}
\label{defterm}
\end{definition}

\paragraph{Syntactic trees.} To any term there is associated a syntactic tree, which is a planar binary tree with nodes decorated with $\lambda$ or with elements $\varepsilon \in \Gamma$ and leaves decorated by terms. According to the definition of terms, any node decorated by $\lambda$ has its left peg decorated with a leaf which is a variable.

\paragraph{Variables, free and bound.}

The functions $\displaystyle Var: T \rightarrow 2^{X}$ and $\displaystyle FV: T \rightarrow 2^{X}$ associate to any term the set of its variables and of its free variables, respectively. These functions are defined according to the following rules. 

\begin{enumerate}
\item[-] if $x \in X$ then $Var(x) = \left\{ x \right\}$ and $FV(x) =  \left\{ x \right\}$
\item[-] if $x \in X$ and $A \in T$ then $Var(x \lambda A) = Var(A) \cup \left\{ x \right\}$ and $FV(x \lambda A)  = FV(A) \setminus \left\{ x \right\}$, 
\item[-] if $A, B \in T$ and $\varepsilon \in \Gamma$ then $Var(A \varepsilon B) = Var(A) \cup Var(B)$ and $FV(A \varepsilon B) = FV(A) \cup FV(B)$
\end{enumerate}

Bound variables are those which appear in a term in the left hand side of an abstraction operation.

\paragraph{Notations in $\lambda$-calculus compared with those in $\lambda$-Scale-calculus.} 
If $x \in X$ is a variable and $A$ is a term then $x \lambda A$ is a term. The corresponding term in  $\lambda$-calculus is: 
$$\lambda x . A = x \lambda A$$

As an example, to the combinator $K = \lambda x. (\lambda y . x)$ from $\lambda$-calculus, corresponds the term $K = x \lambda (y \lambda x)$ from $\lambda$-Scale-calculus.

If $A$ and $B$ are terms, then in $\lambda$ -calculus we have the application operation which sends the pair $(A,B)$ to the term $AB$. Here, in $\lambda$-Scale-calculus,  we shall define the application from the other operations. 

As an example, it is not obvious how to define the combinator $S = \lambda xyz . ((xz)(yz))$ from $\lambda$-calculus. There is though a resemblance between the syntactic tree of $Sxyz$ and the tree associated to the difference operation from emergent algebras $\displaystyle \Delta^{z}(y,x)$, illustrated in the next figure. 

\centerline{\includegraphics[width=100mm]{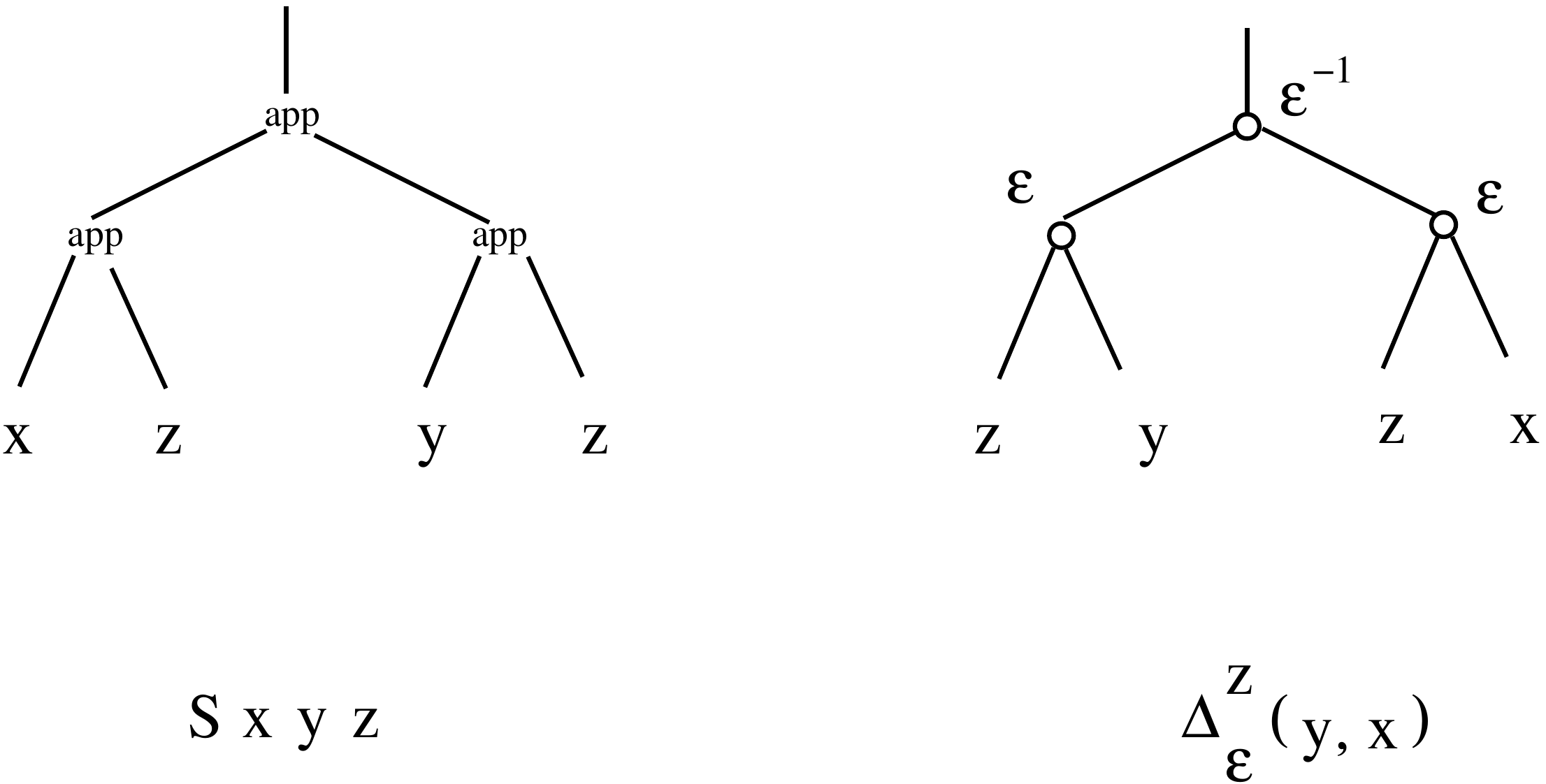}} 

This figure suggests to think about the $\varepsilon$ operation in $\lambda$-Scale-calculus as if it is the following composition between the $\displaystyle \circ_{\varepsilon}$ operation from emergent algebras and the $app$ application from $\lambda$-calculus: 

\centerline{\includegraphics[width=80mm]{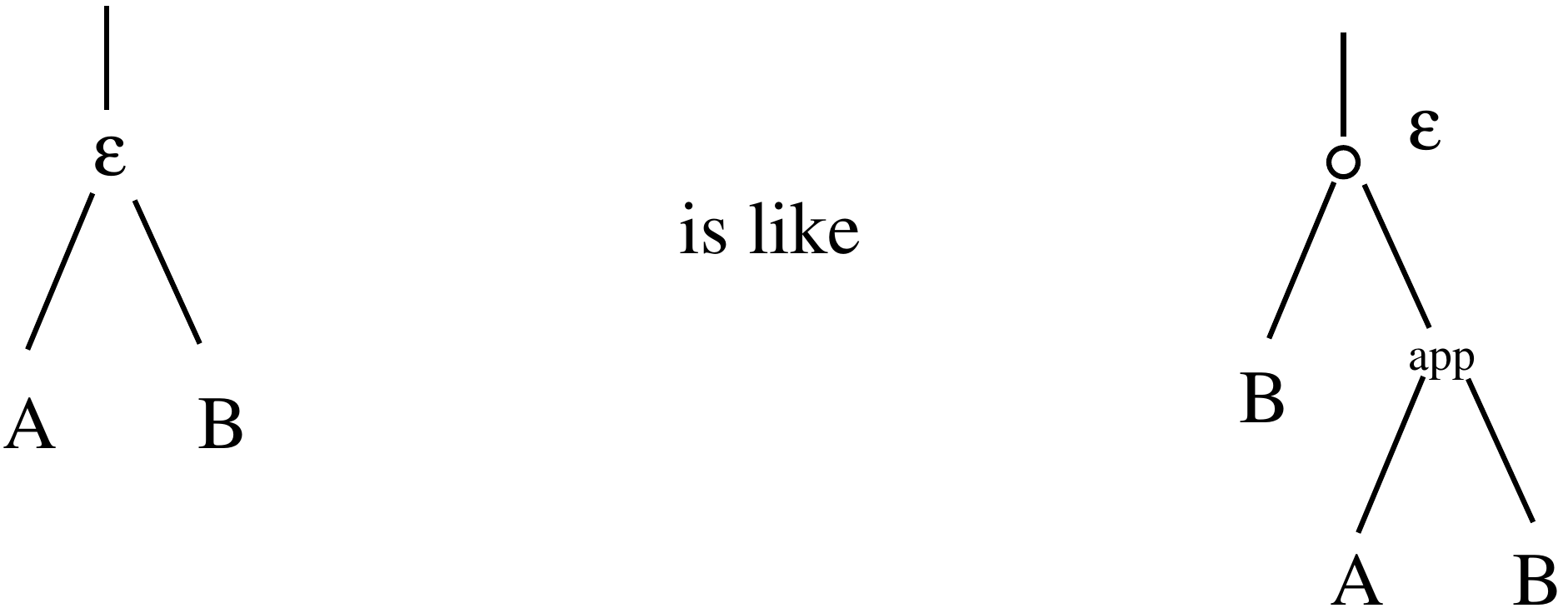}} 

According to the rules of $\lambda$-calculus, we then have 

\vspace{.4cm}

\centerline{\includegraphics[width=100mm]{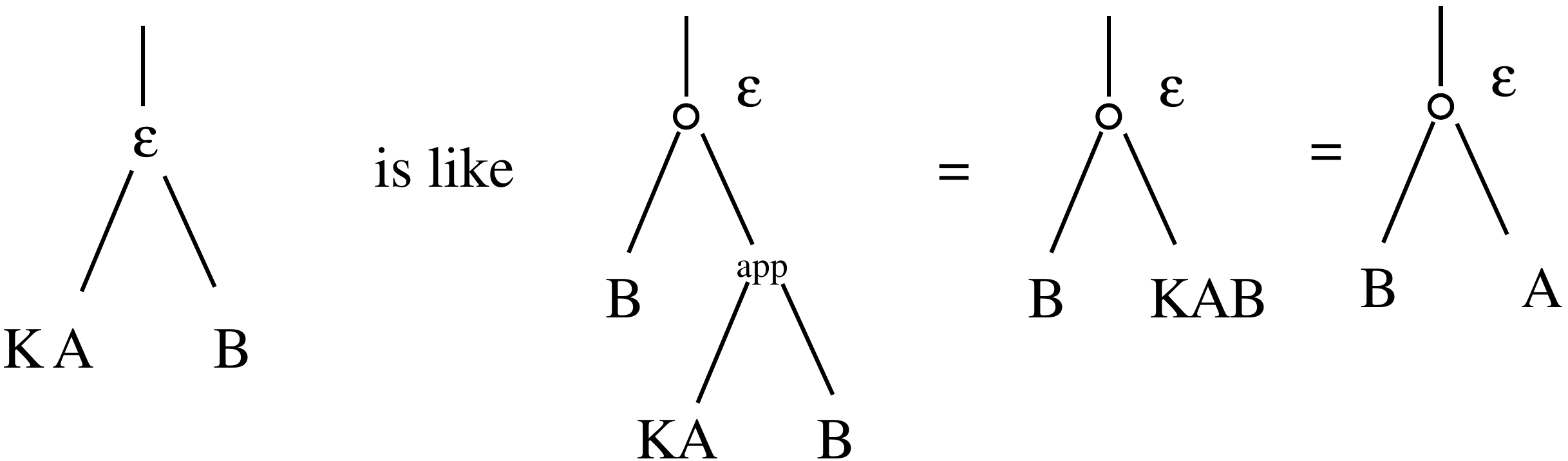}} 

But in $\lambda$-calculus $KA = \lambda y.A$, which we may interpret in $\lambda$-Scale-calculus as $y \lambda A$ (with $y \not \in FV(A)$). 

This leads us to the following definition of the "dilation operation" $\displaystyle \circ_{\varepsilon}$ in $\lambda$-Scale-calculus. 

\begin{definition}
(Dilations) If $A, B$ are terms and $y$ is a variable which does not belong to $FV(A)$ and $\varepsilon \in \Gamma$ then $\displaystyle B\circ_{\varepsilon} A = (y \lambda A)\varepsilon B$. 
\label{defcirceps}
\end{definition}

The definition may be expressed with syntactic trees. 

\centerline{\includegraphics[width=100mm]{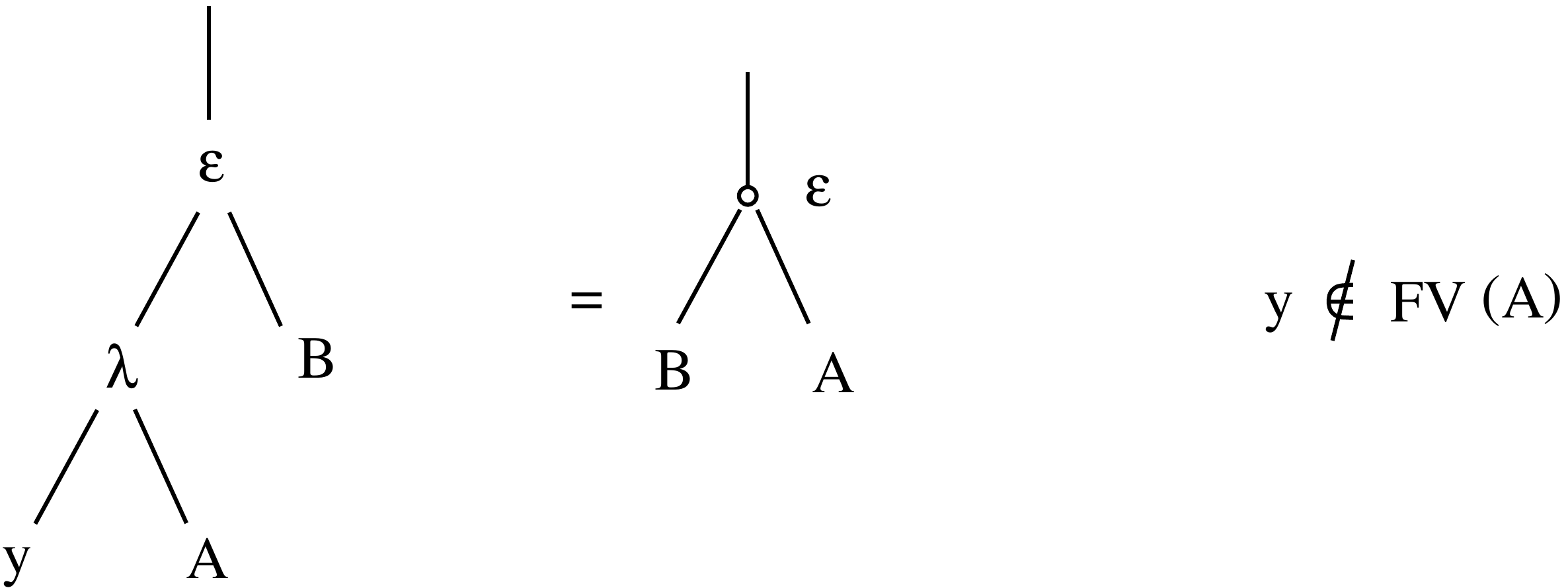}}

\subsection{Reduction and substitution}
\label{secred}

On the set of terms $T$ we shal consider the equivalence relation $\equiv$, which is the transitive ans symmetric closure of the reunion of smaller relations called: $\alpha$-conversion, $\beta$-reduction, the R-moves and ext-reduction. 

The relation $\equiv$ has the following properties. 

\begin{definition}
On the set of terms $T$ we put an equivalence relation $\equiv$, such that for any $\varepsilon \in \Gamma$, any $A, B, C \in T$ and any $x \in X$, $A \equiv B$ implies $A\varepsilon C \equiv B\varepsilon C$ and $B\varepsilon C \equiv A\varepsilon C$ and $(x \lambda A) \equiv (x \lambda B)$. 
\label{defequiv}
\end{definition}

\paragraph{$\alpha$-conversion,} or $\alpha$-renaming, which allows bound variables to be renamed, works as in the usual $\lambda$-calculus. Two terms are  equivalent if one is obtained from the other by an $\alpha$-conversion.

\paragraph{Substitution.} Substitution is the process of replacing all free occurences of a variable $v$ in a term $A$  by a term $B$. The notation is $A[v := B]$. 

\begin{definition}
Substitution is defined  according to the rules: 
\begin{enumerate}
\item[(s1)] if $x \in X$ and $B \in T$ then $x[x:=B]  = B$, 
\item[(s2)] if $x, y \in X$ and $x \not = y$ then $x[y:=B] = x$, 
\item[(s3)] if $A,C \in T$ and $\varepsilon \in \Gamma$ then $(A \varepsilon C) [x :=B] = 
(A[x:=B]) \varepsilon (C[x:=B])$
\item[(s4)] if  $x \not = y$ and $x \not \in FV(B)$ then $(x \lambda A)[y:=B] = x \lambda (A[y:=B])$, 
\end{enumerate}
\label{defsubst}
\end{definition}

In order for the substitution to work properly, $\alpha$ conversion may be needed, in order to have the names of all bound variables different from the names of free variables. 

\paragraph{$\beta$-reduction.} This rule takes the following form: 

\begin{enumerate}
\item[($\beta$*)] if  $y \not \in FV(B) \cup FV(A[x:=B])$ then 
$\displaystyle (x \lambda A) \varepsilon B \equiv (y \lambda (A[x:=B]))\varepsilon B$
\end{enumerate}

\centerline{\includegraphics[width=100mm]{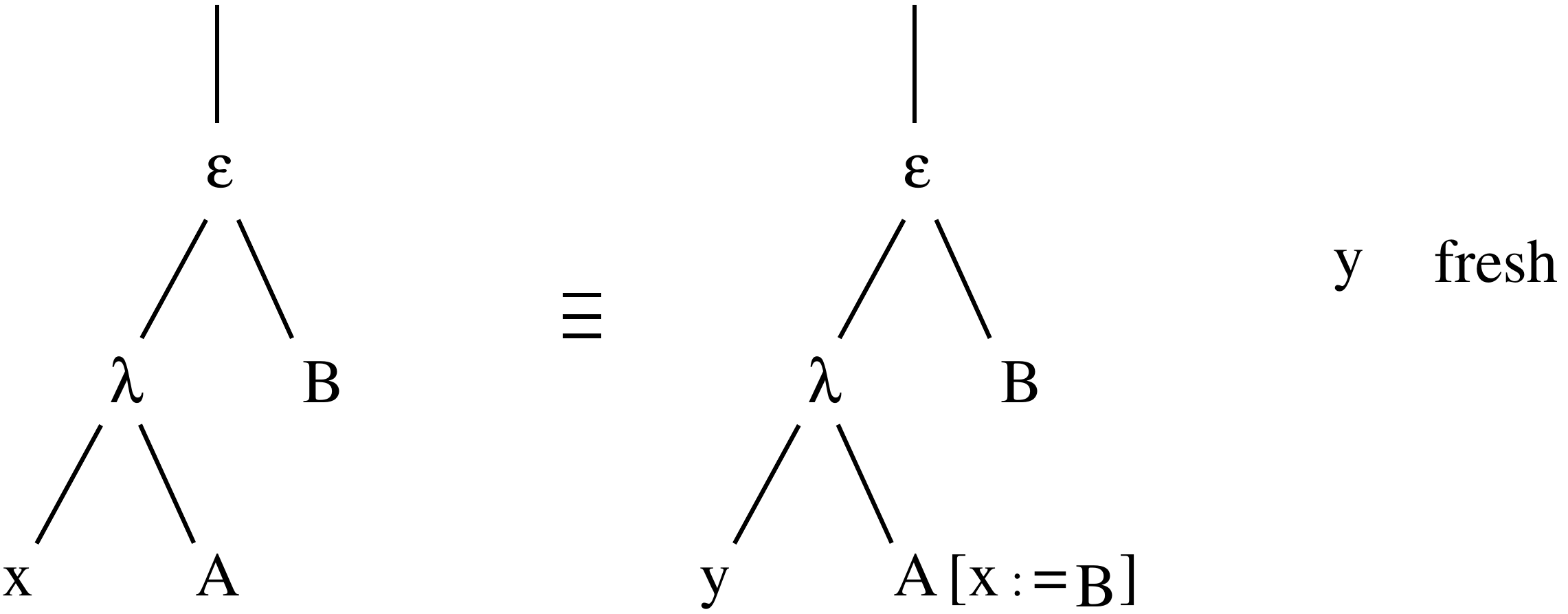}} 

\paragraph{The R moves.} We use definition \ref{defcirceps} in order to import the axioms of $\Gamma$-idempotent right quasigroups into the $\lambda$-Scale-calculus.

\begin{definition}
The two Reidemeister moves in $\lambda$-Scale-calculus are: 
\begin{enumerate}
\item[(R1)] if $x \not \in FV(A)$ then for any $\varepsilon \in \Gamma$ 
$(x\lambda A) \varepsilon A \equiv A$, 

\vspace{.4cm}

\centerline{\includegraphics[width=80mm]{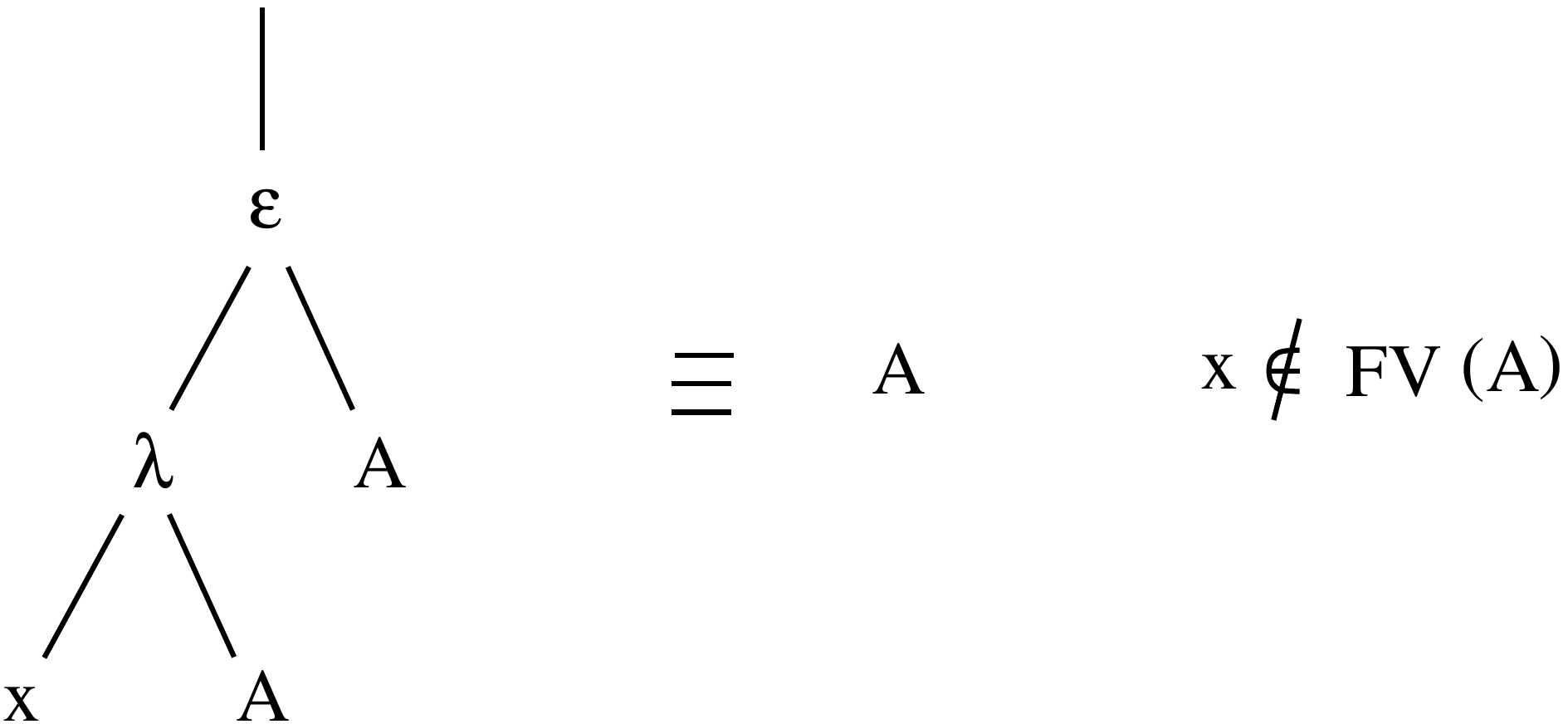}} 

\item[(R2)] if $x \not \in FV(B)$ and $\varepsilon, \mu \in \Gamma$ then 
$(x \lambda (B \mu x)) \varepsilon A \equiv B (\varepsilon \mu) A$.

\vspace{.4cm}

\centerline{\includegraphics[width=80mm]{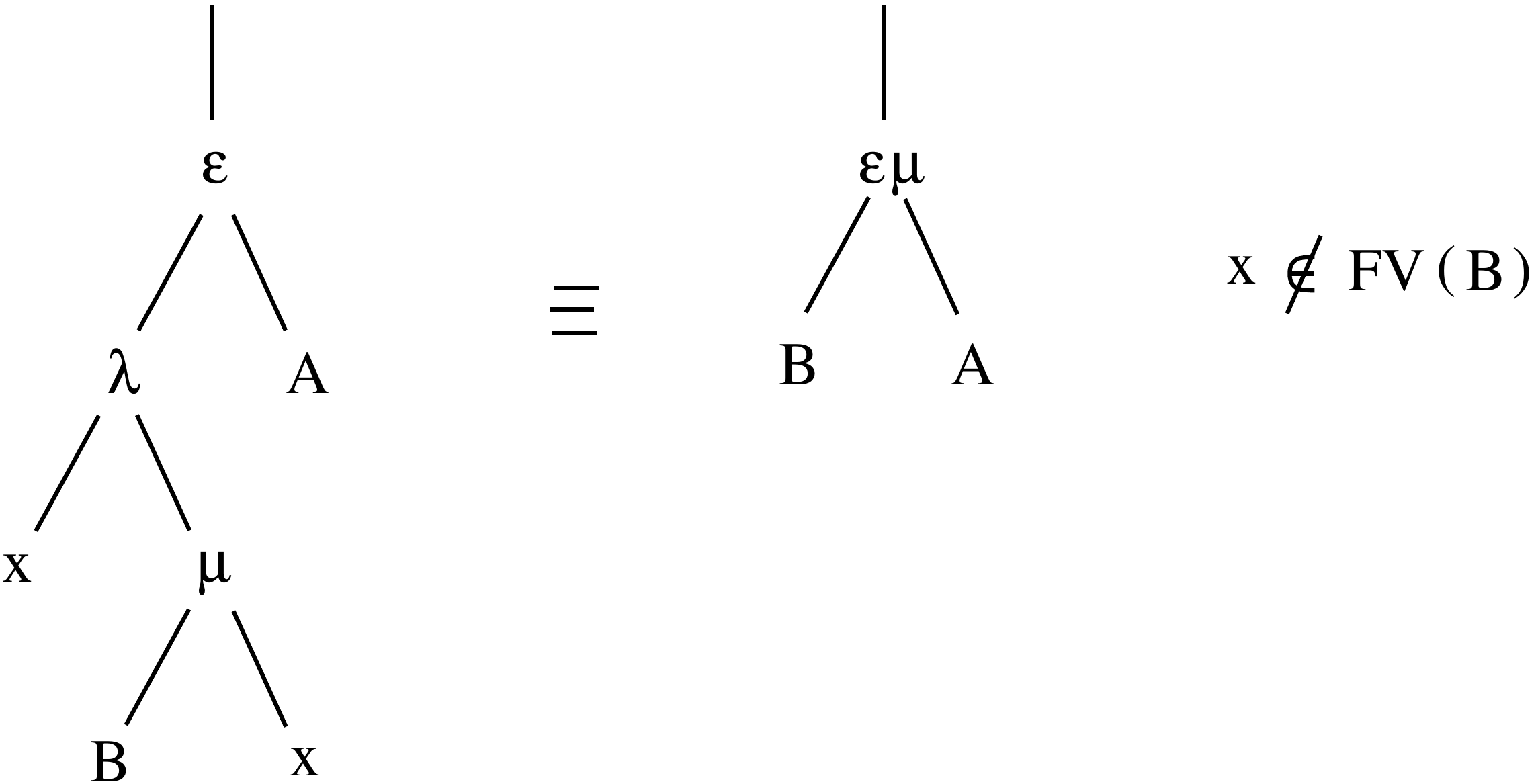}} 

\end{enumerate}
\label{defreid}
\end{definition}

\paragraph{Extensionality rules}
The neutral element of the group $\Gamma$ is denoted by $1$. The following rule (ext1) is the usual  extensionality rule. The second rule (ext2) is imported from emergent algebras. 

\begin{definition}
The extensionality rules are: 
\begin{enumerate}
\item[(ext1)] if $x \not \in FV(B)$ then 
$x \lambda (B 1 x) \equiv B$, 

\vspace{.4cm}

\centerline{\includegraphics[width=80mm]{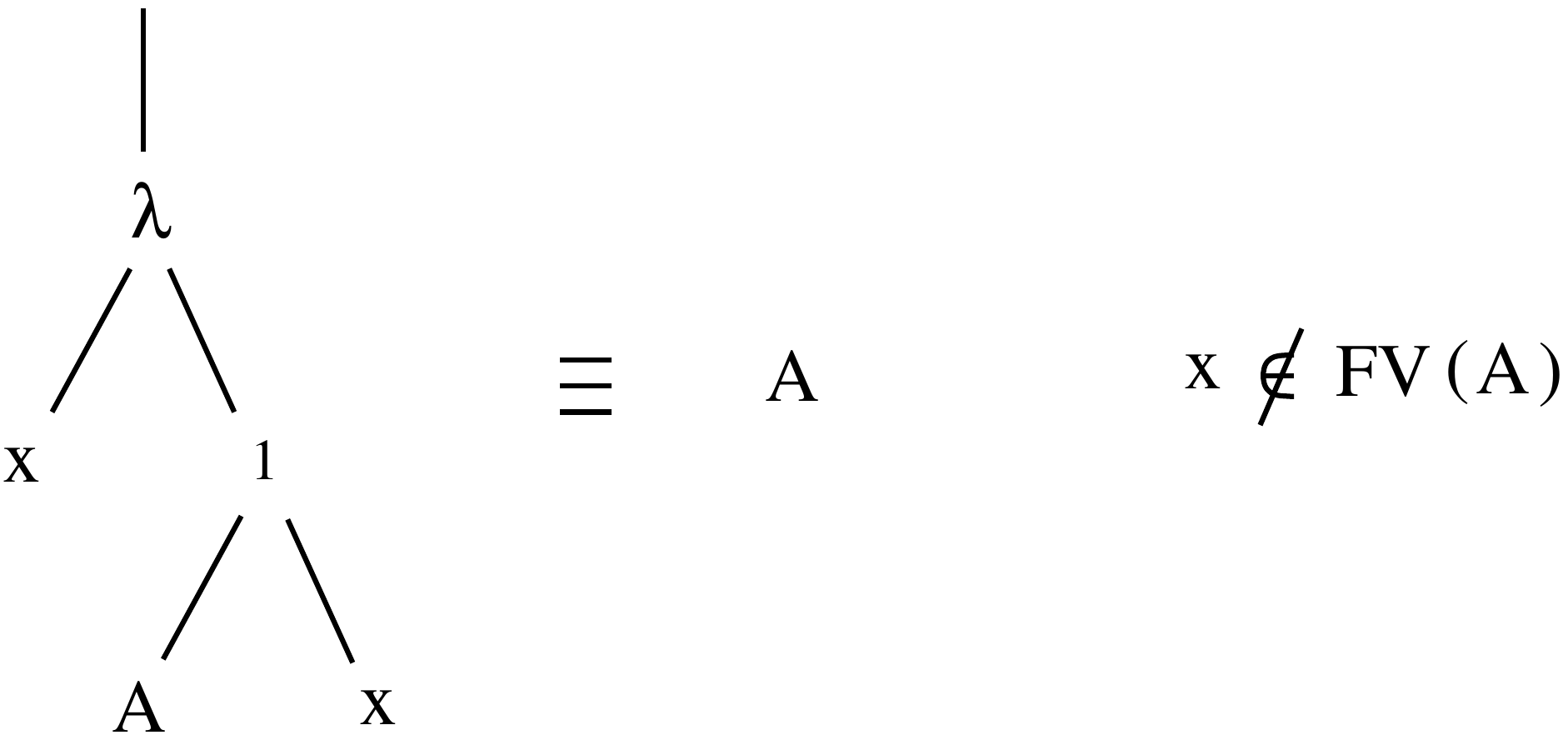}} 

\item[(ext2)] if $x \not \in  FV(B)$ then $(x \lambda B) 1 A = B$.  

\vspace{.4cm}

\centerline{\includegraphics[width=80mm]{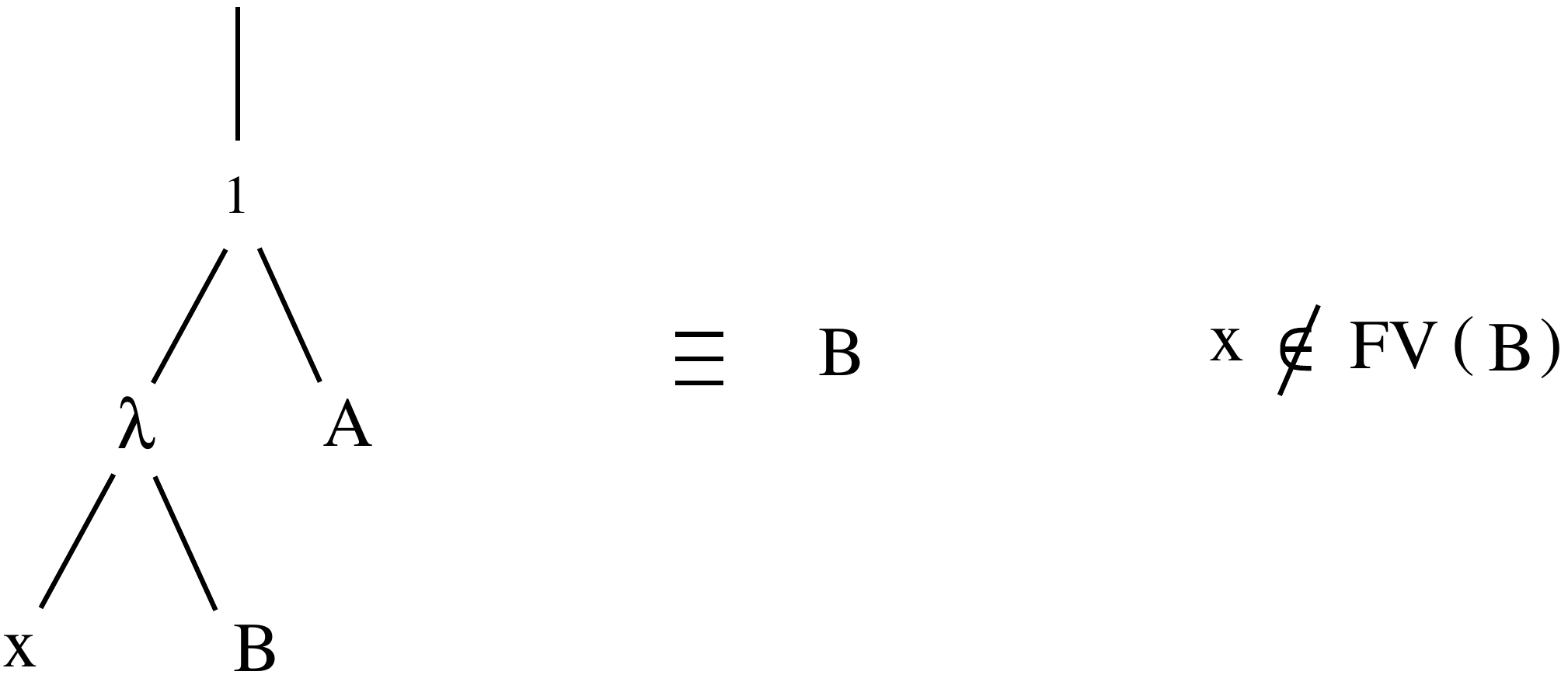}} 

\end{enumerate}
\label{defext}
\end{definition}

\begin{definition}
$\lambda$-Scale calculus is a list $(X,T, \Gamma, comp , abs, \equiv, subst)$, where: 
\begin{enumerate}
\item[-] $\displaystyle comp: T \times  \Gamma  \times T \rightarrow T$ is the composition  function $comp(A, \varepsilon, B) = A \varepsilon B$, 
\item[-]  $abs: X \times T \rightarrow T$ is the abstraction function $abs(x,A) = x \lambda A$, 
\item[-] $T$ is the set of terms constructed from the set of variables $X$, according to definition \ref{defterm}, 
\item[-] $subst$ is the substitution defined  according to definition \ref{defsubst}, rules (s1), (s2), (s3), (s4), 
\item[-] $\equiv$ satisfies definition \ref{defequiv} and ($\beta$*), Reidemeister moves (R1), (R2) and extensionality rules (ext1), (ext2). 
\end{enumerate}
\label{defle}
\end{definition}

\section{$\lambda$-Scale calculus as both lambda calculus and idempotent right quasigroup}

The $app$ operation from $\lambda$-calculus can be defined in the $\lambda$-Scale-calculus as in the following definition. 

\begin{definition}
For any terms $A, B$ we denote by $AB$ the term $A 1 B$. 
\label{defapp}
\end{definition}

\begin{proposition}
For any $\varepsilon \in \Gamma$ and for any terms $A, B$ we have 
$$B \varepsilon A \equiv A \circ_{\varepsilon} (B A)$$
\label{pro1}
\end{proposition}

\centerline{\includegraphics[width=80mm]{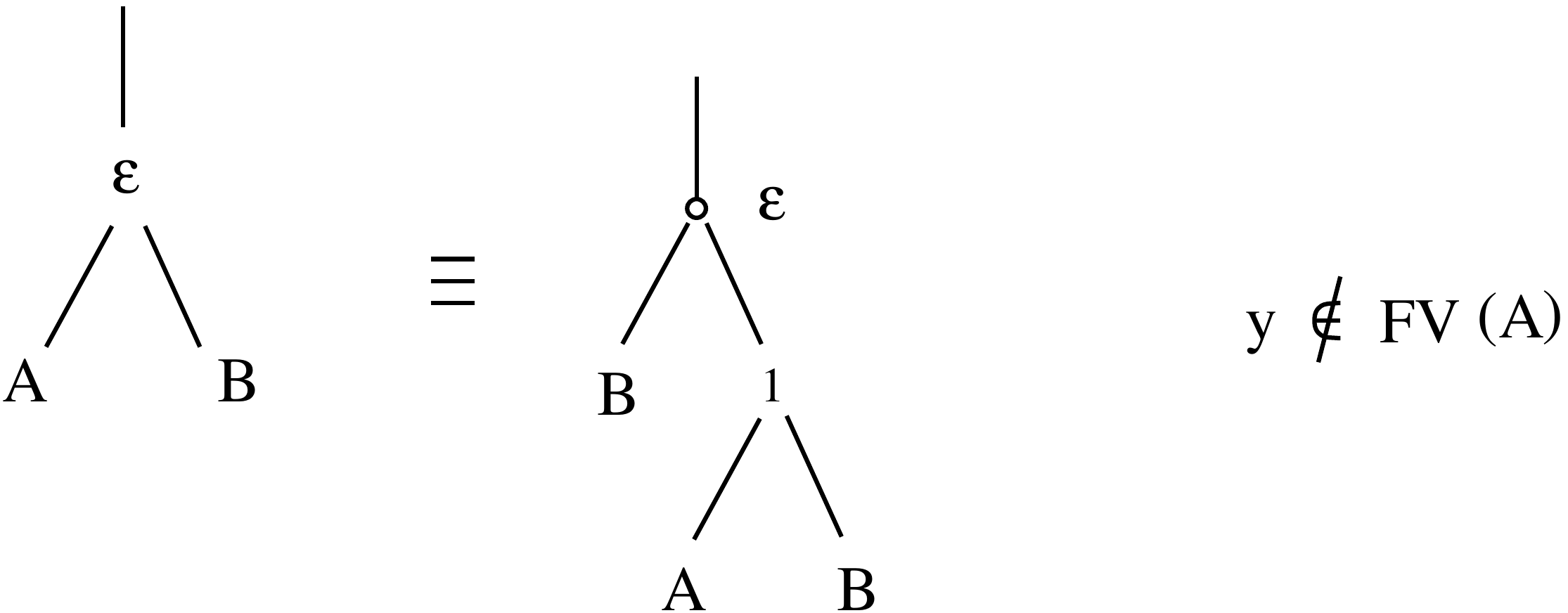}} 

\paragraph{Proof.} By (ext1), if $x \not \in FV(B)$ then $x \lambda ((B 1 x) \equiv B$, therefore $B\varepsilon A \equiv 
(x \lambda ((B 1 x)) \varepsilon A$. By ($\beta$*) we then have 
$$(x \lambda ((B 1 x)) \varepsilon A \equiv (x \lambda ((B 1 A)) \varepsilon A$$
From definition \ref{defcirceps} we get 
$$(x \lambda ((B 1 A)) \varepsilon A = A \circ_{\varepsilon} (B 1 A)$$
therefore we obtain $B \varepsilon A \equiv A \circ_{\varepsilon} (B 1 A)$. \hfill $\square$

Another way of understanding the application operation is provided by the following proposition. The advantage of this "interpretation" of the application operation is that 
it is not using the neutral element of $\Gamma$.

\begin{proposition}
For any terms $A, B$, for any $x \not \in FV(B)$ and for any $\varepsilon \in \Gamma$ we have 
$$ B A \equiv (x\lambda (B (\varepsilon^{-1})x)) \varepsilon A \equiv (x\lambda (B (\varepsilon^{-1})A)) \varepsilon A$$ 
\label{pro2}
\end{proposition}

\paragraph{Proof.}By (R2) and definition \ref{defapp} we have 
$$(x\lambda (B (\varepsilon^{-1})x)) \varepsilon A \equiv  B (\varepsilon \varepsilon^{-1}) A \equiv B1A = B A$$ 
By ($\beta$*) we have 
$$(x\lambda (B (\varepsilon^{-1})x)) \varepsilon A \equiv (x\lambda (B (\varepsilon^{-1})A)) \varepsilon A \quad \square$$

\begin{theorem}
The operation $\displaystyle \circ_{\varepsilon}$ defined up to the equivalence $\equiv$ over  the set $\displaystyle T\diagup  \equiv$, of terms up to equivalence, gives to $\displaystyle T \diagup  \equiv$ the structure of a $\Gamma$-idempotent right quasigroup. 
\label{propirq}
\end{theorem}

\paragraph{Proof.} 
By definition \ref{defequiv}, it is sufficient to prove the content of axioms of a 
$\Gamma$-irq for terms, up to the equivalence $\equiv$. 
The rule (R1) definition \ref{defreid} gives the axiom (R1) from the definition of an irq. 
Let us apply the rule ($\beta$*) for  $x, y \not \in FV(B)$ and $z \not \in FV(A) \cup FV(B)$  and $\varepsilon, \mu \in \Gamma$: 
\begin{equation} 
(x \lambda ((y \lambda B)\mu x )) \varepsilon A \equiv (z \lambda ((y \lambda B)\mu A )) \varepsilon A 
\label{equa1}
\end{equation}
By definition of $\displaystyle \circ_{\varepsilon}, \circ_{\mu}$, the RHS of (\ref{equa1}) is 
$$ (z \lambda ((y \lambda B)\mu A )) \varepsilon A \equiv A \circ_{\varepsilon} ( A \circ_{\mu} B)$$

By rule  (R2) definition \ref{defreid} and then definition of $\displaystyle \circ_{\varepsilon \mu}$, the LHS of (\ref{equa1}) is 
$$(x \lambda ((y \lambda B)\mu x )) \varepsilon A \equiv (y \lambda B) (\varepsilon \mu) A = A \circ_{\varepsilon \mu} B$$
All in all we get: 
$$A \circ_{\varepsilon} ( A \circ_{\mu} B) \equiv A \circ_{\varepsilon \mu} B$$
which, up to equivalence,  is  the axiom (c) from the definition of a $\Gamma$-irq. This, together with the rule (ext2), gives both (R2) from the definition of the irq with the operation $\displaystyle \circ_{\varepsilon}$ (for a fixed $\varepsilon$) and the axiom (b) from the definition of  a $\Gamma$-irq. Thus all is proved. \hfill $\square$

\begin{theorem}
Let us consider the set of terms $\displaystyle T_{1} \subset T$ constructed according to the following rules: 
\begin{enumerate}
\item[(a)] variables $x \in X$ are in  $\displaystyle T_{1}$, 
\item[(b)] if $A, B$ are in  $\displaystyle T_{1}$ then $A B$ is in  $\displaystyle T_{1}$, where 
$A B$ has the meaning from definition \ref{defapp}, 
\item[(c)] if $x$ is a variable and $A$ is a term in  $\displaystyle T_{1}$ then $\lambda x. A$ is in  $\displaystyle T_{1}$, where by definition $\lambda x. A = x \lambda A$, 
\item[(d)] any term in  $\displaystyle T_{1}$ is constructed from a finite number of applications of the previous rules. 
\end{enumerate}
Then  $\displaystyle T_{1}$ with: 
\begin{enumerate}
\item[-] the operations of $\lambda$-abstraction from (c) and application from (b), 
\item[-] together with the substitution rules from definition \ref{defsubst} applied for 
$\varepsilon = 1$ and 
\item[-] the $\alpha$-conversion (applied for terms in  $\displaystyle T_{1}$) and extensionality rule (ext1) 
\end{enumerate}
forms a $\lambda$-calculus. 
\label{propcomb}
\end{theorem}

\paragraph{Proof.}
There is nothing else to mention about $\alpha$-conversion, but the fact that it transforms terms in  $\displaystyle T_{1}$ into terms in  $\displaystyle T_{1}$. The substitution rules (s1)-(s4) from definition \ref{defsubst} transforms into the usual substitution rules for $\lambda$-calculus. It is straightforward  to check that each rule has the property that giving as inputs terms in  $\displaystyle T_{1}$, one gets as output a term in  $\displaystyle T_{1}$.  The rule ($\beta$*) becomes:  if $x \in FV(A)$ and $y \not \in FV(B) \cup FV(A[x:=B])$ then 
$$(\lambda x . A) B \equiv (\lambda y.  (A[x:=B])) B$$ 
which is a rule equivalent to the $\beta$-reduction, using also extensionality (available, see further) and $\alpha$ conversion. 

The rule (ext1) is the usual $\eta$-conversion. The rule (R2) becomes: if $x \not \in FV(B)$ then $(\lambda x. (Bx)) A \equiv BA$, which is a consequence of the $\eta$-conversion.
The rule (R1) becomes: if $x \not \in FV(A)$ then  $(\lambda x. A) A \equiv A$, which is a particular case of the rule (ext2), which takes the form: if $x \not \in FV(B)$ then  $(\lambda x. B) A \equiv B$, which, together with $\eta$-conversion, are extensionality axioms.\hfill $\square$

\section{Relative scaled calculus}
\label{secrel}

Elements  $\varepsilon \in \Gamma$ should be viewed as representing scale. In the following I define $\lambda \epsilon$ calculus at a scale (although this view makes sense only when we contemplate simultaneously all scales). This  is in line with the definition of "chora", section 5 \cite{buligachora}.

In  $\lambda$-Scale calculus we have three operations (which are not independent), namely the lambda abstraction, the application and the emergent algebra  (one parameter family of) operation(s), called dilations. If we want to obtain a scaled version then we have to "conjugate" with dilations. Looking at terms as being syntactic trees, this amounts to:

\begin{enumerate}
\item[-] start with a term $A$ and a scale $\varepsilon \in \Gamma$,
\item[-] transform a term $B$ such that $FV(B) \cap FV(A) = \emptyset$,  into another term  $\displaystyle A_{\varepsilon}[B]$, by conjugating with $\displaystyle A \circ_{\varepsilon} \cdot$.
\end{enumerate}

This can be done by recursively defining the transform $\displaystyle B \mapsto A_{\varepsilon}[B]$. Graphically, we would like to transform the elementary syntactic trees of the three operations into this:

\centerline{\includegraphics[width=80mm]{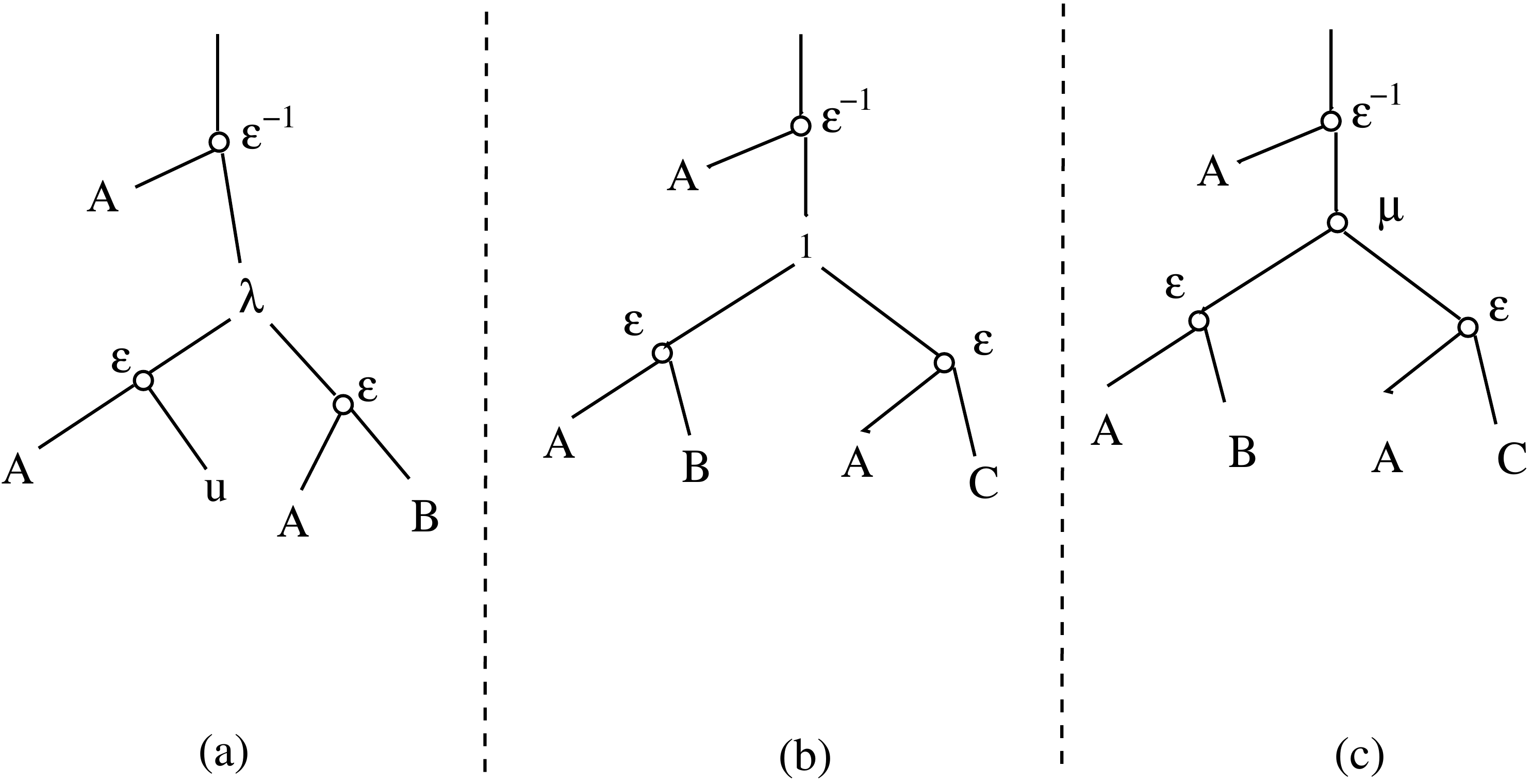}}

The problem is that, while (c) is just the familiar scaled dilation, the scaled $\lambda$ from (a) does not make sense, because $\displaystyle A \circ_{\varepsilon} u$ is not a variable. Also, the scaled application (b) is somehow misterious.

The solution is to exploit the fact that it is possible (although not yet rigorously defined  in this calculus) to  make substitutions of the form 
$\displaystyle B[ A \circ_{\varepsilon} u : = C]$ because of the invertibility of dilations. 
Indeed we may solve the equation  $\displaystyle A \circ_{\varepsilon} u = C$ to get  $\displaystyle u = A \circ_{\varepsilon^{-1}} C$, therefore we may define $\displaystyle B[ A \circ_{\varepsilon} u : = C]$ to mean $\displaystyle B[u : = A \circ_{\varepsilon^{-1}} C]$.

Let us use this in the context of the rule (ext2): consider $B, C$ which have no free variables in common with the ones of $A$. The expression $\displaystyle (A \circ_{\varepsilon} u) \lambda B) 1 C$ should then be equal to $\displaystyle B[ A \circ_{\varepsilon} u : = C]$, that is to 
$\displaystyle B[u : = A \circ_{\varepsilon^{-1}} C] = (u \lambda B) 1 (A \circ_{\varepsilon^{-1}} C)$. 

Graphically, this can be condensed into this figure: 

\centerline{\includegraphics[width=80mm]{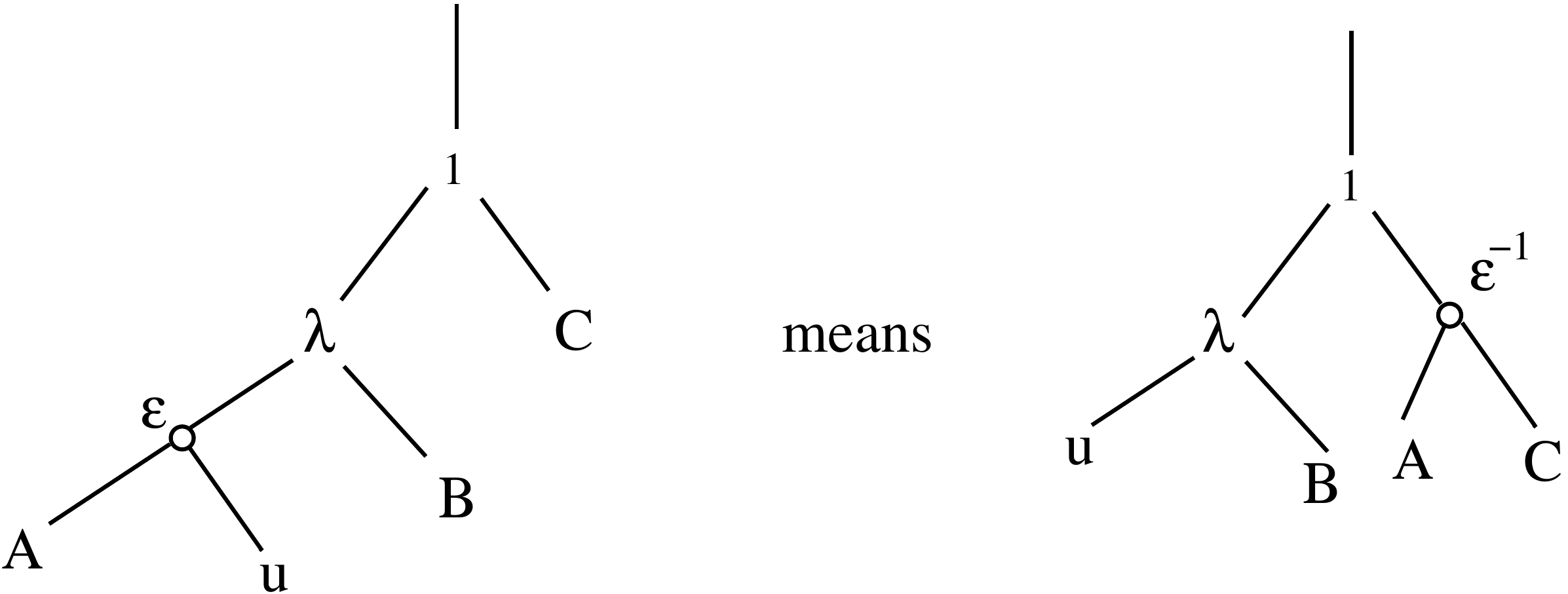}}

But there is a well defined term $T$ with the property that $\displaystyle T 1 C \equiv B[u : = A \circ_{\varepsilon^{-1}} C]$ for any term $C$. Indeed, take 
$$T = z \lambda ( ( u \lambda B) 1 (A \circ_{\varepsilon^{-1}} z))$$ 
Then $\displaystyle T 1 C = ( z \lambda ( ( u \lambda B) 1 (A \circ_{\varepsilon^{-1}} z))) 1 C$ which by (ext2) becomes $T 1 C \equiv ( u \lambda B) 1 (A \circ_{\varepsilon^{-1}} C) \equiv B[u : = A \circ_{\varepsilon^{-1}} C]$. Therefore the correctly defined term which corresponds to the scaled $\lambda$ abstraction should be the one described in the next figure. 

\centerline{\includegraphics[width=80mm]{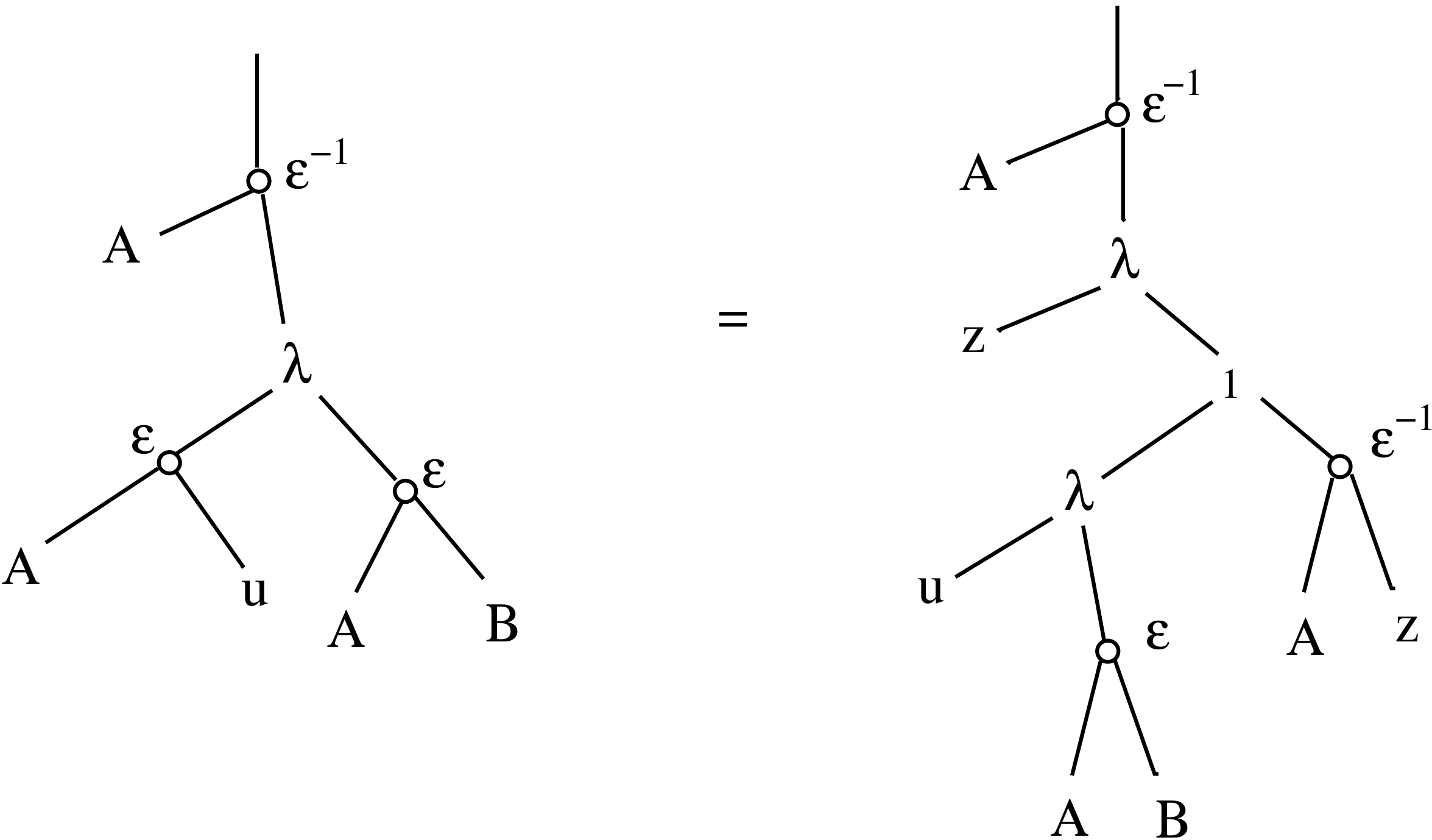}}
(the syntactic tree in the LHS should be seen as a notation for the term in the RHS). 

Let us now start with the construction of the scaled $\lambda$-Scale calculus. 

\begin{definition}
Let us consider $\varepsilon \in \Gamma$ and a term $A \in T\diagup  \equiv$.  Any $\displaystyle  u \in X \setminus FV(A)$ defines a relative variable $\displaystyle  u^{\varepsilon}_{A} := A \circ_{\varepsilon} u$ (remark that relative variables are terms up to equivalence "$\equiv$").The set of relative variables is denoted by $\displaystyle X^{\varepsilon}_{A}$.   

The term $B$ is a scaled term, or relative term, if it belongs to the set $\displaystyle  T^{\varepsilon}_{A}$, defined by: 
\begin{enumerate}
\item[-] scaled variables are relative terms, $\displaystyle X^{\varepsilon}_{A} \subset T^{\varepsilon}_{A}$, 
\item[-] for any $\mu \in \Gamma$ and $\displaystyle B, C \in T^{\varepsilon}_A$ the scaled application (of coefficient $\mu$), $\displaystyle B \mu^{\varepsilon}_{A} C$ is a relative term, 
\item[-] for any scaled variable  $\displaystyle u^{\varepsilon}_{A} \in X^{\varepsilon}_{A}$  and any scaled term $\displaystyle B \in T^{\varepsilon}_{A}$ the scaled abstraction $\displaystyle u^{\varepsilon}_{A} \lambda^{\varepsilon}_{A} B$ is a relative term, 
\item[-] any relative term is obtained after a finite combination of the previous rules. 
\end{enumerate}
\label{deftermeps}
\end{definition}

A relative term has a syntactic tree with respect to the relative operations and relative variables. Let us call this syntactic tree the relative syntactic tree of the relative term.  The relative free variables function poses no problem, being defined with respect to this relative syntactic tree. 

The $\alpha$-conversion or renaming works well on the relative variables. Notice that for any two variables $u,v \in X \setminus FV(A)$ the equality $\displaystyle u^{\varepsilon}_{A} = v^{\varepsilon}_{A}$ is equivalent with $u =v$.

We need a way to translate relative syntactic trees into the initially defined syntactic trees. 
That means we need a function $\displaystyle E^{\varepsilon}_{A}: T^{\varepsilon}_{A} \rightarrow T\diagup \equiv$ which will translate a relative term into a term (up to equivalence "$\equiv$). 
\begin{definition}
The translation function $\displaystyle E^{\varepsilon}_{A}: T^{\varepsilon}_{A} \rightarrow T\diagup \equiv$ is defined inductively on relative terms by: 
\begin{enumerate}
\item[-] $\displaystyle E^{\varepsilon}_{A}[x^{\varepsilon}_{A}] = x$, 
\item[-] $\displaystyle E^{\varepsilon}_{A}[B \mu^{\varepsilon}_{A} C] = A  \circ_{\varepsilon^{-1}} \left( \left( A \circ_{\varepsilon} E^{\varepsilon}_{A}[B] \right) \mu \left(A\circ_{\varepsilon} E^{\varepsilon}_{A}[C]\right) \right)$, 
\item[-] $\displaystyle E^{\varepsilon}_{A}[u^{\varepsilon}_{A} \lambda^{\varepsilon}_{A} B] = A \circ_{\varepsilon^{-1}} (z \lambda (( ( E^{\varepsilon}_{A} [u^{\varepsilon}_{A}] ) \lambda (A \circ_{\varepsilon} E^{\varepsilon}_{A}[B])) 1 ( A \circ_{\varepsilon^{-1}} z)))$
\end{enumerate}
\label{deftranslate}
\end{definition}

This translation function induces a relative equivalence. 

\begin{definition}
Two relative terms $\displaystyle B, C \in T^{\varepsilon}_{A}$ are equivalent, 
notation $\displaystyle B \equiv^{\varepsilon}_{A} C$, if $\displaystyle E^{\varepsilon}_{A}[B] = E^{\varepsilon}_{A}[C]$. 
\label{difrelequiv}
\end{definition}

We shall define the relative substitution like we did in definition \ref{defsubst}. 

\begin{definition}
Relative substitution is defined up to relative $\alpha$-conversion, according to the rules: 
\begin{enumerate}
\item[(rs1)] if $\displaystyle x^{\varepsilon}_{A} \in X^{\varepsilon}_{A}$ and $\displaystyle 
B \in T^{\varepsilon}_{A}$ then $\displaystyle x^{\varepsilon}_{A} [x^{\varepsilon}_{A}:=B]  = B$, 
\item[(rs2)] if $\displaystyle x^{\varepsilon}_{A}, y^{\varepsilon}_{A} \in X^{\varepsilon}_{A}$ are different relative variables then $\displaystyle x^{\varepsilon}_{A} [y^{\varepsilon}_{A}:=B]  =  x^{\varepsilon}_{A}$, 
\item[(rs3)] if $\displaystyle D,C \in T^{\varepsilon}_{A}$ and $\mu \in \Gamma$ then $\displaystyle (D 
\mu^{\varepsilon}_{A} C) [x^{\varepsilon}_{A} :=B] = 
(D[x^{\varepsilon}_{A}:=B]) \mu^{\varepsilon}_{A} (C[x^{\varepsilon}_{A}:=B])$
\item[(rs4)] if  $\displaystyle x^{\varepsilon}_{A} \not = y^{\varepsilon}_{A}$ and 
$\displaystyle x^{\varepsilon}_{A} \not \in FV(B)$ then $(x^{\varepsilon}_{A} \lambda^{\varepsilon}_{A} C)[y^{\varepsilon}_{A}:=B] = x^{\varepsilon}_{A} \lambda^{\varepsilon}_{A} (C[y:=B])$, 
\end{enumerate}
\label{defrelsubst}
\end{definition}

\begin{proposition}
$\displaystyle E^{\varepsilon}_{A} \left[B[x^{\varepsilon}_{A} : = C]\right] = \left(E^{\varepsilon}_{A}\left[B\right] \right)[x: =  E^{\varepsilon}_{A}[C]]$

\label{prelsub}
\end{proposition}

\paragraph{Proof.} By induction on terms. Indeed, if $\displaystyle B = x^{\varepsilon}_{A}$ then $\displaystyle E^{\varepsilon}_{A} \left[B[x^{\varepsilon}_{A} : = C]\right] = E^{\varepsilon}_{A} \left[x^{\varepsilon}_{A}[x^{\varepsilon}_{A} : = C]\right] = E^{\varepsilon}_{A} \left[C\right]$ and $\displaystyle \left(E^{\varepsilon}_{A}\left[B\right] \right)[x: = E^{\varepsilon}_{A}[C]] = E^{\varepsilon}_{A}[C]$. 

If $\displaystyle B = u^{\varepsilon}_{A} \not = x^{\varepsilon}_{A}$ then $\displaystyle E^{\varepsilon}_{A} \left[B[x^{\varepsilon}_{A} : = C]\right] = E^{\varepsilon}_{A} \left[u^{\varepsilon}_{A}[x^{\varepsilon}_{A} : = C]\right] = E^{\varepsilon}_{A} \left[u^{\varepsilon}_{A}\right] = u$ and $\displaystyle \left(E^{\varepsilon}_{A}\left[u^{\varepsilon}_{A}\right] \right)[x: = E^{\varepsilon}_{A}[C]] = u[x: = E^{\varepsilon}_{A}[C]] = u$.

If $\displaystyle E^{\varepsilon}_{A} \left[B[x^{\varepsilon}_{A} : = C]\right] = \left(E^{\varepsilon}_{A}\left[B\right] \right)[x: =  E^{\varepsilon}_{A}[C]]$ and 
$\displaystyle E^{\varepsilon}_{A} \left[D[x^{\varepsilon}_{A} : = C]\right] = \left(E^{\varepsilon}_{A}\left[D\right] \right)[x: =  E^{\varepsilon}_{A}[C]]$ then 
$\displaystyle E^{\varepsilon}_{A} \left[\left(B \mu^{\varepsilon}_{A} D \right)[x^{\varepsilon}_{A} : = C]\right] = \left(E^{\varepsilon}_{A}\left[\left(B \mu^{\varepsilon}_{A} D \right)\right] \right)[x: =  E^{\varepsilon}_{A}[C]]$.

Finally, if $\displaystyle E^{\varepsilon}_{A} \left[B[x^{\varepsilon}_{A} : = C]\right] = \left(E^{\varepsilon}_{A}\left[B\right] \right)[x: =  E^{\varepsilon}_{A}[C]]$ then then for any $u \not = x$ we have 
$$ E^{\varepsilon}_{A} \left[u^{\varepsilon}_{A} \lambda^{\varepsilon}_{A} B\right] [x: =  E^{\varepsilon}_{A}[C]] = \left( A \circ_{\varepsilon^{-1}} (z \lambda ((  u \lambda (A \circ_{\varepsilon} E^{\varepsilon}_{A}[B])) 1 ( A \circ_{\varepsilon^{-1}} z))) \right) [x: =  E^{\varepsilon}_{A}[C]]$$
We use the hypothesis and (rs4) to obtain: 
$$ E^{\varepsilon}_{A} \left[u^{\varepsilon}_{A} \lambda^{\varepsilon}_{A} B\right] [x: =  E^{\varepsilon}_{A}[C]] = E^{\varepsilon}_{A} \left[ u^{\varepsilon}_{A} \lambda^{\varepsilon}_{A} \left( B[x^{\varepsilon}_{A} : = C] \right) \right] = E^{\varepsilon}_{A}\left[ \left(u^{\varepsilon}_{A} \lambda^{\varepsilon}_{A} B\right)[x^{\varepsilon}_{A} : = C]\right]$$
The proof is done. \hfill $\square$

We give a simpler form of the  translation of $\displaystyle u^{\varepsilon}_{A} \lambda^{\varepsilon}_{A} B$ in the next proposition. 

\begin{proposition}
Let $\displaystyle u^{\varepsilon}_{A} \in X^{\varepsilon}_{A}$ and $\displaystyle B \in T^{\varepsilon}_{A}$. Then 
$$\displaystyle E^{\varepsilon}_{A} \left[ u^{\varepsilon}_{A} \lambda^{\varepsilon}_{A} B \right]  = 
A \circ_{\varepsilon^{-1}} (z \lambda (A \circ_{\varepsilon} \left( \left(E^{\varepsilon}_{A}[B]\right)[u := A\circ_{\varepsilon^{-1}} z] \right)  ))$$
\label{psimply}
\end{proposition}

\paragraph{Proof.} Indeed, $\displaystyle E^{\varepsilon}_{A} \left[ u^{\varepsilon}_{A} \lambda^{\varepsilon}_{A} B \right] = A \circ_{\varepsilon^{-1}} (z \lambda ((  u \lambda (A \circ_{\varepsilon} E^{\varepsilon}_{A}[B])) 1 ( A \circ_{\varepsilon^{-1}} z)))$. By 
($\beta$*), for variable $u'$ fresh,  $\displaystyle E^{\varepsilon}_{A} \left[ u^{\varepsilon}_{A} \lambda^{\varepsilon}_{A} B \right] = A \circ_{\varepsilon^{-1}} (z \lambda ( (u' \lambda (A \circ_{\varepsilon} \left(\left(E^{\varepsilon}_{A}[B]\right)[u := A\circ_{\varepsilon^{-1}} z ]\right))) 1 ( A \circ_{\varepsilon^{-1}} z )))$.  By (ext2) we get 
$\displaystyle E^{\varepsilon}_{A} \left[ u^{\varepsilon}_{A} \lambda^{\varepsilon}_{A} B \right] = A \circ_{\varepsilon^{-1}} (z \lambda (A \circ_{\varepsilon} \left( \left(E^{\varepsilon}_{A}[B]\right)[u := A\circ_{\varepsilon^{-1}} z\right) ]  ))$. \hfill $\square$

\begin{theorem}
The list $\displaystyle (X^{\varepsilon}_{A}, T^{\varepsilon}_{A}, \Gamma, comp^{\varepsilon}_{A}, abs^{\varepsilon}_{A}, \equiv^{\varepsilon}_{A} , relsubst)$, where: 
\begin{enumerate}
\item[-] $\displaystyle comp^{\varepsilon}_{A}: T^{\varepsilon}_{A} \times  \Gamma  \times T^{\varepsilon}_{A} \rightarrow T^{\varepsilon}_{A}$ is the composition  function $comp^{\varepsilon}_{A}(B, \mu, C) = B \mu^{\varepsilon}_{A} C$, 
\item[-]  $\displaystyle abs^{\varepsilon}_{A}: X^{\varepsilon}_{A} \times T^{\varepsilon}_{A} \rightarrow T^{\varepsilon}_{A}$ is the abstraction function $\displaystyle abs^{\varepsilon}_{A}(x,B) = x^{\varepsilon}_{A} \lambda^{\varepsilon}_{A} B$, 
\item[-] $relsubst$ is the relative substitution, definition \ref{defrelsubst}, 
\end{enumerate}
is a $\lambda$-Scale calculus according to definition \ref{defle}. 
\label{pscaledcalc}
\end{theorem}

\paragraph{Proof.}  
We have to prove ($\beta$*), (R1), (R2), (ext1), (ext2). 

($\beta$*): let $\displaystyle x^{\varepsilon}_{A} \in FV(C)$. Then:  
$$\displaystyle 
E^{\varepsilon}_{A} \left[ \left( x^{\varepsilon}_{A} \lambda^{\varepsilon}_{A} C \right) \mu^{\varepsilon}_{A} B\right] = A \circ_{\varepsilon^{-1}} \left(\left( z \lambda 
\left( A \circ_{\varepsilon} \left( \left( E^{\varepsilon}_{A}[C]\right)\left[x:= A \circ_{\varepsilon^{-1}} z \right]  \right)\right) \right)\mu \left(A \circ_{\varepsilon} E^{\varepsilon}_{A}\left[ B \right] \right) \right) $$ by proposition \ref{psimply}. By ($\beta$*) and $\displaystyle A \circ_{\varepsilon^{-1}}  \left( A \circ_{\varepsilon}  E^{\varepsilon}_{A}[B] \right) =  E^{\varepsilon}_{A}[B]$ the term from the RHS simplifies to 
\begin{equation}
A \circ_{\varepsilon^{-1}} \left( \left( y \lambda \left( A \circ_{\varepsilon} \left( \left( E^{\varepsilon}_{A}[C]\right)\left[ x:= E^{\varepsilon}_{A}[B]\right] \right)\right)\right) \mu \left( A \circ_{\varepsilon} E^{\varepsilon}_{A}[B] \right) \right)
\label{middle}
\end{equation}
with $y$ a fresh variable. Now we compute: 
$$\displaystyle E^{\varepsilon}_{A} \left[ \left( y^{\varepsilon}_{A} \lambda^{\varepsilon}_{A} \left(C[x^{\varepsilon}_{A}:= B] \right) \right) \mu^{\varepsilon}_{A} B\right] = $$ 
$$ = A \circ_{\varepsilon^{-1}} \left(\left( z \lambda 
\left( A \circ_{\varepsilon} \left( \left( E^{\varepsilon}_{A}[C[x^{\varepsilon}_{A}:= B]]\right) \right)\right) \right)\mu \left(A \circ_{\varepsilon} E^{\varepsilon}_{A}\left[ B \right] \right) \right)$$
by proposition \ref{psimply} and the fact that $y$ is a fresh variable. We use now proposition \ref{prelsub} and we reduce further the RHS of this equality to the term (\ref{middle}). All in all we proved: 
$$\displaystyle 
E^{\varepsilon}_{A} \left[ \left( x^{\varepsilon}_{A} \lambda^{\varepsilon}_{A} C \right) \mu^{\varepsilon}_{A} B\right] = E^{\varepsilon}_{A} \left[ \left( y^{\varepsilon}_{A} \lambda^{\varepsilon}_{A} \left(C[x^{\varepsilon}_{A}:= B] \right) \right) \mu^{\varepsilon}_{A} B\right]$$ which is equivalent to 
$$\left( x^{\varepsilon}_{A} \lambda^{\varepsilon}_{A} C \right) \mu^{\varepsilon}_{A} \equiv^{\varepsilon}_{A}  \left( y^{\varepsilon}_{A} \lambda^{\varepsilon}_{A} \left(C[x^{\varepsilon}_{A}:= B] \right) \right) \mu^{\varepsilon}_{A} B $$

(R1): if $\displaystyle x^{\varepsilon}_{A} \not \in FV(B)$ then 
by proposition \ref{psimply}, by $\displaystyle A \circ_{\varepsilon^{-1}}  \left( A \circ_{\varepsilon}  D \right) =  D$, for any term $D \in T\diagup \equiv$, by definition of dilations \ref{defcirceps} we get 
$$ E^{\varepsilon}_{A} \left[ \left( x^{\varepsilon}_{A} \lambda^{\varepsilon}_{A} B \right) \mu^{\varepsilon}_{A} B\right] = A \circ_{\varepsilon^{-1}} \left(  \left( A \circ_{\varepsilon} E^{\varepsilon}_{A}[B]\right) \circ_{\mu} \left( A \circ_{\varepsilon} E^{\varepsilon}_{A}[B] \right)     \right)$$ 
We apply the Reidemeister move (R1) (for dilations) to the RHS and obtain 
$$ E^{\varepsilon}_{A} \left[ \left( x^{\varepsilon}_{A} \lambda^{\varepsilon}_{A} B \right) \mu^{\varepsilon}_{A} B\right] = A \circ_{\varepsilon^{-1}} \left( A \circ_{\varepsilon}  E^{\varepsilon}_{A}[B] \right) = E^{\varepsilon}_{A}[B]$$ 
which proves that $\displaystyle \left( x^{\varepsilon}_{A} \lambda^{\varepsilon}_{A} B \right) \mu^{\varepsilon}_{A} B  \equiv^{\varepsilon}_{A} B$. 

(R2): a similar proof, only that we use (R2) for dilations instead of (R1). 

(ext1): We use proposition \ref{psimply} to write that, for $\displaystyle x^{\varepsilon}_{A} \not \in FV(B)$,  
$$E^{\varepsilon}_{A} \left[ x^{\varepsilon}_{A} \lambda^{\varepsilon}_{A} \left( B  1^{\varepsilon}_{A} x^{\varepsilon}_{A}  \right)\right] = 
A \circ_{\varepsilon^{-1}} \left( z \lambda \left( A \circ_{\varepsilon} 
\left( E^{\varepsilon}_{A}\left[ B  1^{\varepsilon}_{A} x^{\varepsilon}_{A} \right] \left[ x:=  A \circ_{\varepsilon^{-1}} z  \right] \right) \right) \right) = $$
$$= A \circ_{\varepsilon^{-1}} \left( z \lambda \left( 
\left( \left(A \circ_{\varepsilon} E^{\varepsilon}_{A}\left[ B \right] \right)  1 \left( A \circ_{\varepsilon} x \right)\right) 
\left[ x:=  A \circ_{\varepsilon^{-1}} z  \right]  \right) \right) =$$ 
$$ =  A \circ_{\varepsilon^{-1}} \left( z \lambda \left(     \left(A \circ_{\varepsilon} E^{\varepsilon}_{A}\left[ B \right] \right)  1 z   \right) \right)$$
By (ext1) applied to the last term we obtain: 
$$E^{\varepsilon}_{A} \left[ x^{\varepsilon}_{A} \lambda^{\varepsilon}_{A} \left( B  1^{\varepsilon}_{A} x^{\varepsilon}_{A}  \right)\right] = A \circ_{\varepsilon^{-1}} \left( 
A \circ_{\varepsilon} E^{\varepsilon}_{A}\left[ B \right] \right) = E^{\varepsilon}_{A}\left[ B \right]$$
The last relation (ext2) has a similar proof. \hfill $\square$


\begin{thebibliography}{99}


\bibitem{baren} H.P. Barendregt, The Lambda Calculus: Its Syntax and Semantics, Studies in Logic and the Foundations of Mathematics, 103 (Revised ed.), North Holland, Amsterdam

\bibitem{buligadil1} M. Buliga, Dilatation structures I. Fundamentals, {\it 
J. Gen. Lie Theory Appl.},  {\bf 1} (2007),  2, 65-95. 



\bibitem{buligairq} M. Buliga, Emergent algebras, \url{http://arxiv.org/abs/0907.1520}

\bibitem{buligabraided} M. Buliga, Braided spaces with dilations and sub-riemannian symmetric spaces, in: Geometry. Exploratory Workshop on Differential Geometry and its Applications, eds. D. Andrica, S. Moroianu, Cluj-Napoca 2011, 21-35,  
\url{http://arxiv.org/abs/1005.5031}

\bibitem{buligachora} M. Buliga, Computing with space: a tangle formalism for chora and difference , \url{http://arxiv.org/abs/1103.6007}


\bibitem{buligaintro} M, Buliga, Introduction to metric spaces with dilations
(2010), \url{http://arxiv.org/abs/1007.2362}



\bibitem{church} A. Church, A set of postulates for the foundation of logic, Annals of Mathematics, Series 2, 33:346–366 (1932)




\bibitem{fennrourke} R. Fenn, C. Rourke, Racks and Links in codimension two, 
 {\it J. Knot Theory Ramifications},  {\bf 1}  (1992),  no. 4, 343--406



\bibitem{joyce} D. Joyce, A classifying invariant of knots; the knot quandle, 
{\it J. Pure Appl. Alg.}, {\bf 23} (1982), 37-65 



\bibitem{vodosel} S.K. Vodopyanov, S.V. Selivanova, Algebraic properties of the tangent cone to a quasimetric space with dilations, {\bf Doklady Math.} 2 (2009), 734-738


\end{thebibliography}
\end{document}